%% file: Tau96_preprint.tex
\documentstyle[twoside,fleqn,espcrc2]{article}

\input myphyx

\input{psfig}

\newcommand{\tauegamma}{\tau^- \ra e^-\gamma}
\newcommand{\taumugamma}{\tau^- \ra \mu^-\gamma}
\newcommand{\mueee}{\mu^- \ra e^-e^+e^-\nuebar\nu_\mu}
\newcommand{\taueee}{\tau^- \ra   e^-e^+e^-\bar \nu_e   \nutau}
\newcommand{\taumuee}{\tau^- \ra \mu^-e^+e^-\bar \nu_\mu \nutau}
\newcommand{\tauemumu}{\tau^- \ra e^-\mu^+\mu^-\bar \nu_e \nutau}
\newcommand{\taumumumu}{\tau^- \ra \mu^-\mu^+\mu^-\bar \nu_\mu \nutau}
\newcommand{\taulll}{\tau^- \ra l^-l^+l^-\bar \nu_l \nutau}

\newcommand{\AmS}{{\protect\the\textfont2
  A\kern-.1667em\lower.5ex\hbox{M}\kern-.125emS}}

\title{Review of Rare and Forbidden $\tau$ Decays}

\author{K.K. GAN\address{Department of Physics,
        The Ohio State University,
        Columbus, OH 43210,
        U.S.A.}
        \thanks{Invited talk presented at the Fourth International
                Workshop on Tau Lepton Physcis, Estes Park, Colorado,
                16-19 September, 1996}}
       
\begin{document}

\begin{abstract}
This is a review of rare and forbidden decays of the $\tau$ lepton.
For the rare decays, this includes new results on the chiral anomaly
decay $\taupietapio$, new upper limits on the second-class-current
decay $\taupieta$, and the observations of the Cabibbo-suppressed decay
$\tauketa$ and the internal conversion decay $\taueee$.
For the forbidden decays, there are new upper limits
on the radiative decays $\tauegamma$ and $\taumugamma$.
Some forbidden decays which have not been previously searched for
are also suggested.
\end{abstract}

\maketitle
\section{INTRODUCTION}

\renewcommand{\thepage}{ }
\thispagestyle{myheadings}
\markright{\rm \hfill OHSTPY-HEP-E-96-039}

\vspace{-0.2cm}
Rare and forbidden decays of the $\tau$ lepton are of particular interest.
In rare decays, the Standard Model interaction is suppressed
and therefore the sensitivity to new physics may be enhanced.
In forbidden decays, the observation of a signal would imply
physics beyond the Standard Model.
The $\tau$ lepton is an excellent laboratory for the search
for physics beyond the Standard Model.
Its large mass allows for searches at high momentum transfer with
many decay channels.
The sensitivity may be enhanced because the $\tau$ is a third generation lepton.
In some models, the coupling may have a mass dependence, e.g.
$\propto m_\tau^5$, resulting in higher sensitivity than searches using $\mu$ decay.
In this paper, we first review the rare decays, the chiral anomaly
decay $\taupietapio$, the second-class-current decay $\taupieta$,
the Cabibbo-suppressed decay $\tauketa$,
and the internal conversion decay $\taueee$.
This is followed by the review of forbidden decays, including new
upper limits on the radiative decays $\tauegamma$ and $\taumugamma$.
Some forbidden decays which have not been previously searched for
are also suggested.

\vspace{-0.2cm}
\section{CHIRAL ANOMALY DECAY $\taupietapio$}

\begin{figure}[htbp]
\centerline{\psfig{figure=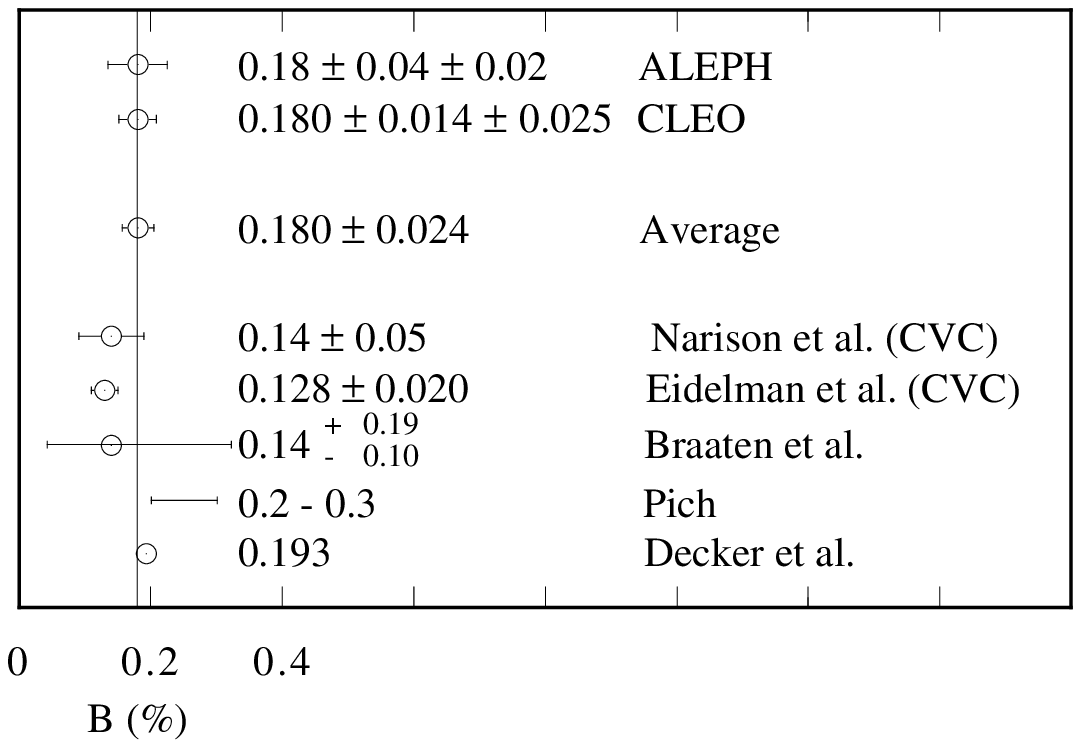,width=3.0in}}
\vspace{-0.5cm}
\caption{
The measured branching ratios for $\tau^- \ra h^-\eta\pi^0\nutau$
and the predictions for $\taupietapio$.}
\label{fig:B_pietapi0}
\end{figure}

\vspace{-0.1cm}
The decay $\taupietapio$ proceeds through the vector current
since the $G$-parity of the $\pi\eta\pi^0$ system is positive.
However, in the chiral limit, the vector current couples
exclusively to even numbers of pseudo-scalars.
The decay is thus a chiral anomaly of QCD;
the decay proceeds through the Wess-Zumino \cite{Wess}
anomaly term in the effective Lagrangian, changing
the parity of three-meson final state and hence permitting
the decay without isospin suppression.
The calculations of the decay rate have large
uncertainties \cite{Braaten,Pich,Decker} as shown in Fig.~\ref{fig:B_pietapi0}.
The rate can be more reliably estimated by using the
conserved-vector-current (CVC) \cite{Feynman} hypothesis to relate
the coupling strength of the three-meson system to the weak charged
vector current and the electromagnetic neutral vector current.
The predictions for the rate \cite{Narison,Eidelman} estimated
using the measured cross section for $\ee \to \pi^+\pi^-\eta$
are shown in Fig.~\ref{fig:B_pietapi0}.
In 1992, the decay $\tau^- \ra h^-\eta\pi^0\nutau$ was observed
for the first time by CLEO~II \cite{CLEO_pietapi0}.
In the measurement, there is no attempt to distinguish the charged
particle as $\pi$ or $K$; the rate for $\tau^- \ra K^-\eta\pi^0\nutau$
is expected to be highly suppressed.
The CLEO experiment has updated (preliminary) its measurement using a five
times larger data sample \cite{Shelkov}.
The decay has now been observed by ALEPH \cite{ALEPH_pietapi0} as well.
The two measurements are consistent with each other
as shown in Fig.~\ref{fig:B_pietapi0}.
The measurements are also consistent with the predictions
based on an effective Lagrangian.
However, they are both somewhat higher than the CVC expectations.
This potential discrepancy may be resolved soon with a higher
precision measurement of the $\pi^+\pi^-\eta$ cross section
expected from Novosibirsk \cite{Eidelman}.

\section{SECOND-CLASS-CURRENT DECAY $\taupieta$}

\thispagestyle{myheadings}
\markright{}

In the Standard Model, the weak hadronic current has a $V-A$ structure
and the hadronic decay products have distinctive charge conjugation
$(C)$ and isospin (and hence $G$-parity) signatures, a reflection
of the quantum number of the charged hadronic weak current.
The weak current is classified according to its $G$-parity:
\begin{eqnarray*}
\hspace{0.6cm} G = -1 \hspace{1cm} J^P = 0^-, 1^+ \hspace{1.0cm} \pi, a_1...\\
\hspace{0.6cm} G = +1 \hspace{1cm} J^P = 1^-      \hspace{1.5cm} \rho...\hspace{0.6cm} \\
\end{eqnarray*}
These are known as the first class currents.  Currents with opposite
$G$-parity are called the second class currents \cite{Weinberg}.
Examples of second class current decays are
$\tau^- \to a^-_0(980)\nutau \to \pi^-\eta\nutau$ and
$\tau^-\ra b^-_1(1235)\nutau \to \pi^-\omega\nutau$.

The classification of the decay $\taupieta$ as a second class current
can be understood by analyzing the $J^P$ of the $\pi\eta$ system:
If system has $J = 0$, then
\begin{eqnarray*}
\hspace{0.3cm}P = P(\pi)P(\eta)(-1)^J = (-1)(-1)(-1)^0 = +1\\
\end{eqnarray*}
and hence $J^P = 0^+$.
On the other hand, if $J = 1$, then
\begin{eqnarray*}
\hspace{0.3cm}P = P(\pi)P(\eta)(-1)^J = (-1)(-1)(-1)^1 = -1,\\
\end{eqnarray*}
which gives $J^P = 1^-$.
However, the $G$-parity of the system is
\begin{eqnarray*}
\hspace{1.0cm}G = G(\pi)G(\eta) = (-1)(+1) = -1,\\
\end{eqnarray*}
which is opposite to that expected for a first class current.
The decay $\taupieta$ is therefore a second class
current regardless of the angular momentum of the $\pi\eta$ system.
The decay is strongly suppressed and there are several theoretical
predictions on the branching ratio \cite{Pich,Tisserant,Neufeld}:
\begin{eqnarray*}
B \sim 1.6 \times 10^{-5} \hspace{0.8cm} \rm Tisserant~and~Truong\hspace{0.85cm}\\
B \sim 1.5 \times 10^{-5} \hspace{0.8cm} \rm Pich\hspace{3.5cm}\\
B \sim 1.2 \times 10^{-5} \hspace{0.8cm} \rm Neufeld~and~Rupertsberger\\
\end{eqnarray*}

\begin{figure}[htbp]
\centerline{\psfig{figure=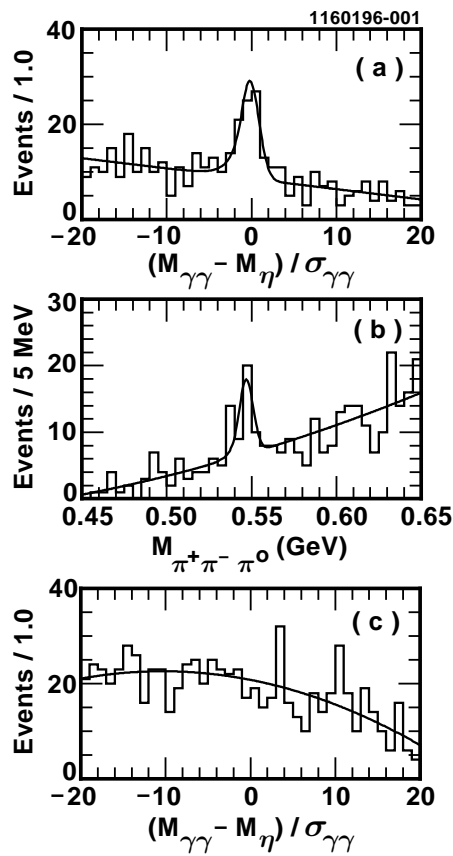,width=2.4in}}
\vspace{-0.5cm}
\caption{The invariant mass spectrum of the $\eta$ candidates.
Each $\eta$ candidate is accompanied by a kaon candidate in
(a) and (b) and by a pion candidate in (c).
Each curve shows a fit to the mass spectrum.
The $\eta$ candidates are selected with an inclusive 1-prong
tag (plus any number of photons) in (a) and (b) and with a
lepton tag in (c).
$\sigma_{\gg}$ is the mass resolution.}
\label{fig:mass_gg}
\end{figure}

The CLEO~II experiment has obtained a new limit on $\taupieta$
in an analysis that also measured the rate for $\tauketa$ \cite{Bartelt1}.
The $\pi$/K identification is based on a confidence level ratio
which is constructed from the confidence levels for
$\pi$ and $K$ hypotheses, ${\cal C}_{\pi}$ and ${\cal C}_K$. 
The confidence level ratio for $\pi$ is
\begin{eqnarray*}
\hspace{1.0cm} R_\pi = \frac{{\cal C}_\pi}{{\cal C}_\pi+{\cal C}_K},
\end{eqnarray*}
and similarly for $K$ ($R_K = 1 - R_\pi$).
The confidence level is computed from the $\chi^2$ probability for a
particle hypothesis using a combination of TOF and dE/dx  information.
A $\pi$ candidate is then defined as  a particle with $R_\pi>0.5$,
otherwise the particle is considered a kaon.
The $\eta$ meson accompanying the $\pi$ candidate is detected via
the decay channel $\eta \to \gg$.
There is no evident for the decay in the $\gg$ invariant mass
spectrum as shown in Fig.~\ref{fig:mass_gg}.
The 95\% confidence level upper limit on the decay is
\begin{eqnarray*}
\hspace{1.0cm} B(\taupieta) < 1.4 \times 10^{-4}.
\end{eqnarray*}
ALEPH has also searched for the decay and the upper
limit \cite{ALEPH_pietapi0} is
\begin{eqnarray*}
\hspace{1.0cm} B(\taupieta) < 6.2 \times 10^{-4}.
\end{eqnarray*}
CLEO's upper limit is about one order of magnitude
above the theoretical expectations \cite{Pich,Tisserant,Neufeld}.

In the Standard Model, the decay $\tau^- \to \pi^-\omega\nutau$
proceeds through the vector current with $J^P = 1^-$ for the
$\pi\omega$ system.
The decay may contain a contribution from the second class current
with $J^P = 0^-$ or $1^+$.
ALEPH has searched for the contribution \cite{ALEPH_pietapi0}
by analyzing the distribution of the $\omega$ decay angle, the angle
between the normal to the $\omega$ decay plane and the direction of
the bachelor pion, in the $\omega$ rest frame.
The upper limit on the second class current contribution
is 8.6\% at the 95\% confidence level.
Using the current world average \cite{PDG} of
$B(\tau^- \to h^-\omega\nutau) = (1.91 \pm 0.09)\%$, this corresponds to
$B(\tau^- \to \pi^-\omega\nutau) < 1.7 \times 10^{-3}$ for
the second-class-current process.

\begin{figure}[htb]
\centerline{\psfig{figure=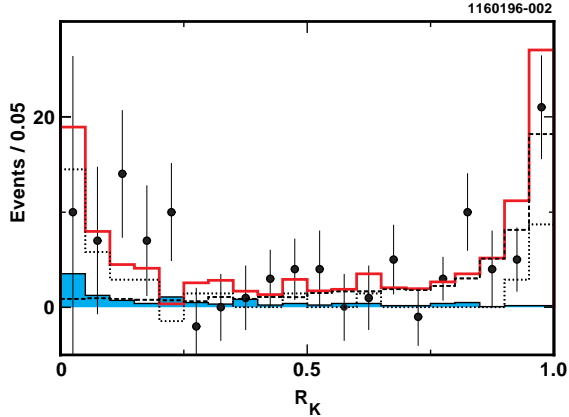,width=3.0in}}
\vspace{-0.5cm}
\caption{The observed $R_K$ spectrum of the charged particle
in the hemisphere containing an $\eta$ meson after
sideband subtraction.
The $\eta$ candidate is selected with an inclusive 1-prong
tag (plus any number of photons).
The histogram shows the Monte Carlo expectation
which is a sum of the predictions for
$\tau^- \to K^- \eta \nu_{\tau}$ (dashed),
$\tau^- \to \pi^- \pi^0 \eta \nu_{\tau}$ (shaded), and
$e^+e^- \to q \bar q$ (dotted).}
\label{fig:R_K}
\end{figure}

It is instructive to assess the prospect for observing the
second-class-current decay $\taupieta$, extrapolated using
the CLEO~II as a prototype detector for the B-factories.
The $R_K$ distribution for the charged track accompanying
the $\eta$ candidate is shown in Fig.~\ref{fig:R_K}.
Candidates for $\taupieta$ populate the low $R_K$ region.
In this region, the largest background is $e^+e^- \to q \bar q$,
followed by $\taupietapio$, and then $\tauketa$.
The hadronic background can be greatly suppressed
by requiring a lepton tag.
Since the CLEO~II calorimeter is close to the state-of-the-art, no major
gain in suppressing the background from $\taupietapio$ is expected.
The $K\eta$ background will be eliminated with the good particle
identification capacity expected for the B-factory detectors.
The signal to noise ratio is therefore expected to be 1:7,
a daunting challenge for the experimenters.

\section{CABIBBO-SUPPRESSED DECAY $\tauketa$}

\begin{figure}[htbp]
\centerline{\psfig{figure=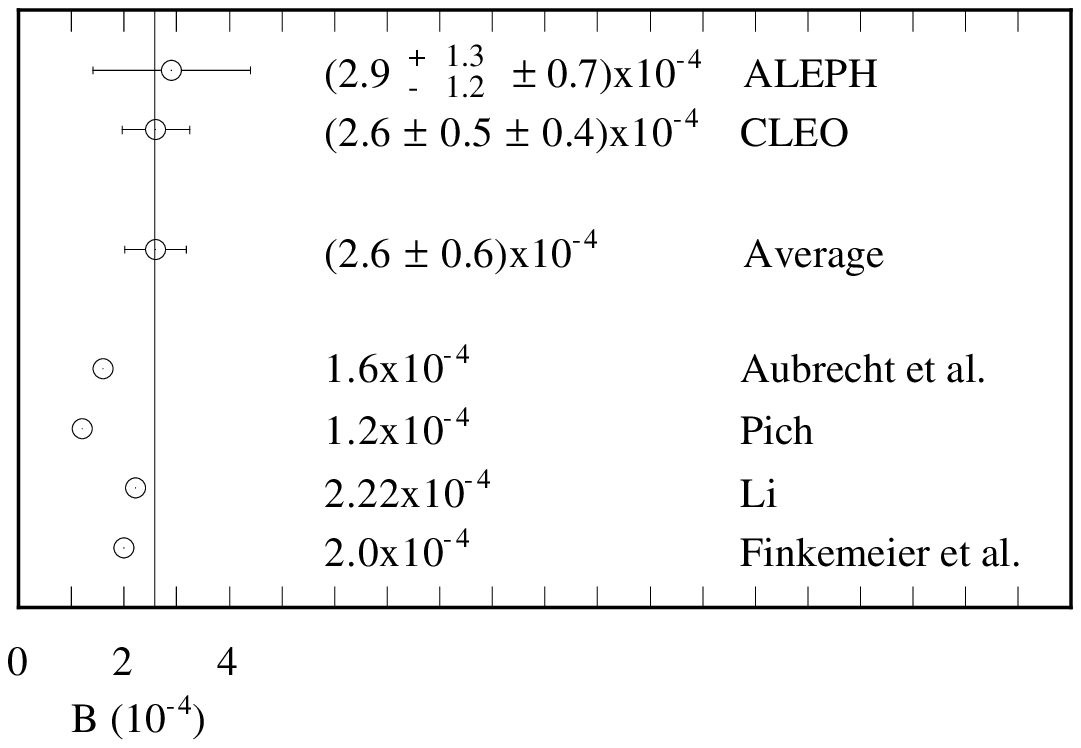,width=3.0in}}
\vspace{-0.5cm}
\caption{
Branching ratios for $\tauketa$.}
\label{fig:B_Keta}
\end{figure}

There is no $G$-parity constraint on the Cabibbo-suppressed decay $\tauketa$,
unlike the decay $\taupieta$, due to SU(3) symmetry breaking.
The branching ratio is expected to be larger than $\taupieta$ by an order of
magnitude \cite{Pich,Aubrecht,Li,Mirkes} as summarized in Fig.~\ref{fig:B_Keta}.

As stated in the previous section, the CLEO~II experiment has
searched for $\tauketa$ in the same analysis that searched for
$\taupieta$ \cite{Bartelt1}.
The $\eta$ meson accompanying the $K$ candidate is detected via
the decay channels $\eta \to \gg$ and $\pi^+\pi^-\pi^0$.
A signal is observed in both channels as shown in Fig.~\ref{fig:mass_gg}.
This is the first evidence for the decay.
CLEO's measurement as well as a recent measurement by
ALEPH \cite{ALEPH_pietapi0} is shown in Fig.~\ref{fig:B_Keta}.
Both results are consistent with each other and
the theoretical expectations \cite{Pich,Aubrecht,Li,Mirkes}.

\section{INTERNAL CONVERSION DECAY $\taulll$}
\begin{figure}[htbp]
\centerline{\psfig{figure=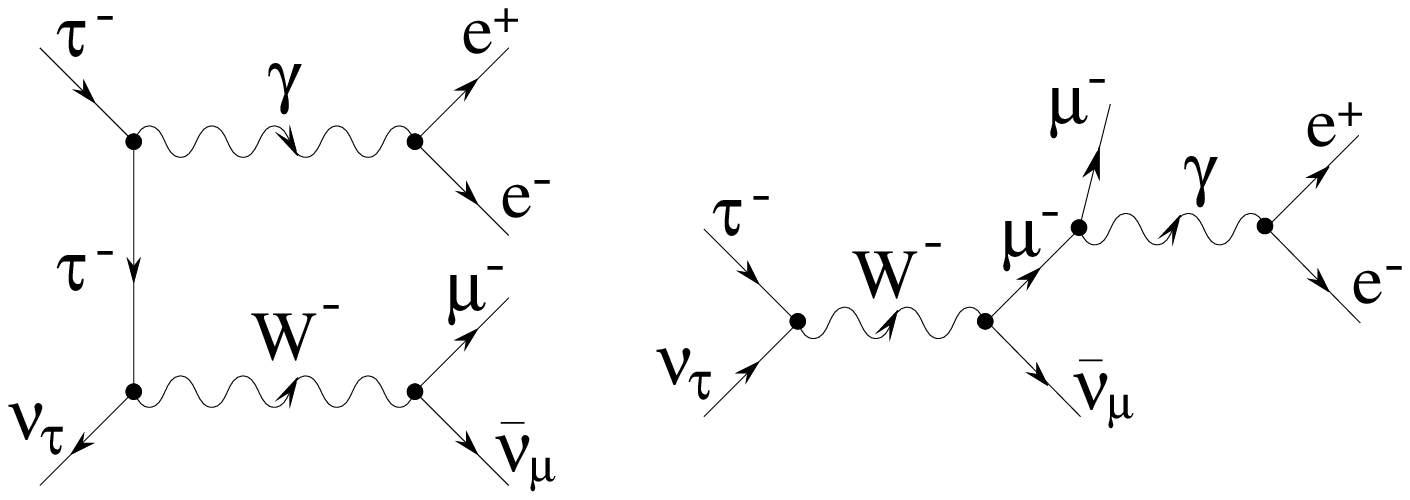,width=3.8in}}
\vspace{-0.5cm}
\caption{
Feynman diagrams for $\taumuee$.}
\label{fig:Feynman}
\end{figure}

The decay of the $\tau$ lepton into three lighter leptons and two
neutrinos is allowed in the Standard Model.
The decay proceeds via the emission of a virtual photon
with subsequent internal conversion into $\ee$
and $\mu^+\mu^-$ as shown in Fig.~\ref{fig:Feynman}.
The contribution from a virtual photon radiated off the
$W$ boson is negligible due to the $W$ propagator.
The branching ratios for various internal conversion decays
have been calculated by Dicus and Vera \cite{Dicus} as
shown in Table~\ref{tab:3l2nu}.
The branching ratios have also been calculated by Volobouev
of the CLEO Collaboration \cite{3l2nu} using
the symbolic manipulation program FORM \cite{FORM}.
These predictions are $\sim$7\% higher, although
they agree on the decay $\mueee$.
An independent calculation would be useful to resolve the difference.
The branching ratios for the decays $\taueee$ and
$\mu^-e^+e^-\bar \nu_\mu \nutau$ are expected to be at
the $10^{-5}$ level and hence within the reach of the CLEO~II experiment.
The internal conversion into a muon pair is suppressed by two
orders of magnitude and hence beyond the reach of the experiment.

The CLEO~II experiment has searched for the decays
$\taueee$ and $\mu^-e^+e^-\bar \nu_\mu \nutau$ \cite{3l2nu}.
Experimentally, this is a very challenging search because
of the difficulty in reconstructing the low momentum tracks from the
internal conversion and special care has been taken in identifying
the soft electron candidates using the specific energy
loss measurements ($dE/dx$). 
Nevertheless, the experiment has identified five candidates for
$\taueee$ and one candidate for $\taumuee$, over a background
of $\sim$0.3 and 0.5 events, respectively.
This yields the branching ratios
\begin{eqnarray*}
\hspace{0.5cm}B(\taueee)\hspace{2.65cm}\\
\hspace{0.5cm}= (2.7^{+1.5}_{-1.1} \pm 0.4 ^{+0.1}_{-0.3})  \times 10^{-5}\hspace{0.5cm}\\
\hspace{0.5cm}B(\taumuee) < 3.2 \times 10^{-5}\hspace{0.5cm}\\
\end{eqnarray*}
at the 95\% confidence level, where the first error is
statistical, the second is systematic, and the third is
due to the uncertainty in the background correction.
The result is consistent with both calculations.
It is remarkable that we are now measure rare decays
at the $10^{-5}$ level.

\begin{table*}[hbt]
\setlength{\tabcolsep}{1.5pc}
\newlength{\digitwidth} \settowidth{\digitwidth}{\rm 0}
\catcode`?=\active \def?{\kern\digitwidth}
\caption{Theoretical predictions on the branching ratios
for $\mu$ and $\tau$ decays with internal conversion.
The errors are due to the inaccuracies in the numerical integrations.}
\label{tab:3l2nu}
\begin{tabular*}{\textwidth}{@{}l@{\extracolsep{\fill}}cc}
\hline
Channel      & Dicus and Vera                   & CLEO Calculation                 \\ \hline
$\taueee$    & $(4.15  \pm 0.06 )\times10^{-5}$ & $(4.457 \pm 0.006)\times10^{-5}$ \\
$\taumuee$   & $(1.97  \pm 0.02 )\times10^{-5}$ & $(2.089 \pm 0.003)\times10^{-5}$ \\
$\tauemumu$  & $(1.257 \pm 0.003)\times10^{-7}$ & $(1.374 \pm 0.002)\times10^{-7}$ \\
$\taumumumu$ & $(1.190 \pm 0.002)\times10^{-7}$ & $(1.276 \pm 0.004)\times10^{-7}$ \\
$\mueee$     & $(3.60  \pm 0.02 )\times10^{-5}$ & $(3.605 \pm 0.005)\times10^{-5}$ \\ \hline
\end{tabular*}
\end{table*}

\section{FORBIDDEN DECAYS}
In the Standard Model, there is no symmetry associated with
lepton flavor and therefore there is no fundamental conservation
law for lepton flavor; lepton flavor conservation is an
experimentally observed phenomenon.
Lepton flavor violation is expected in many extensions of
Standard Model such as lepto-quarks \cite{Pati},
SUSY \cite{Kelley,Barbieri,Hisano,Gomez}, superstrings \cite{Wu},
left-right symmetric \cite{Mophapatra} models and models which
include heavy neutral leptons \cite{Valle,Ilakovoc1,Ilakovoc2}.
The	predictions typically depend on one or two unknown masses
of new particles and one or two unknown couplings.
Therefore any	null result from a search can only constrain the
parameter space but cannot rule out a particular model.
Nevertheless the search should be pursued vigorously because
of its profound implication on the Standard Model should a
positive signal be observed.

The number of $\tau$'s collected or expected for individual
detectors is as follow:

\vspace{2mm}
\hspace{1cm} LEP:\hspace{20mm}$4 \times 10^5~\tau$'s

\vspace{2mm}
\hspace{1cm} CLEO~II:\hspace{19mm}$10^7~\tau$'s

\vspace{2mm}
\hspace{1cm} B-factory:\hspace{18mm}$10^9~\tau$'s

\vspace{2mm}
\noindent So far, all the searches are statistics limited and are not
expected to be background limited in the foreseeable future.
The CLEO~II experiment excels in this kind of search with
the world largest $\tau$ sample.

\begin{figure}[htbp]
\centerline{\psfig{figure=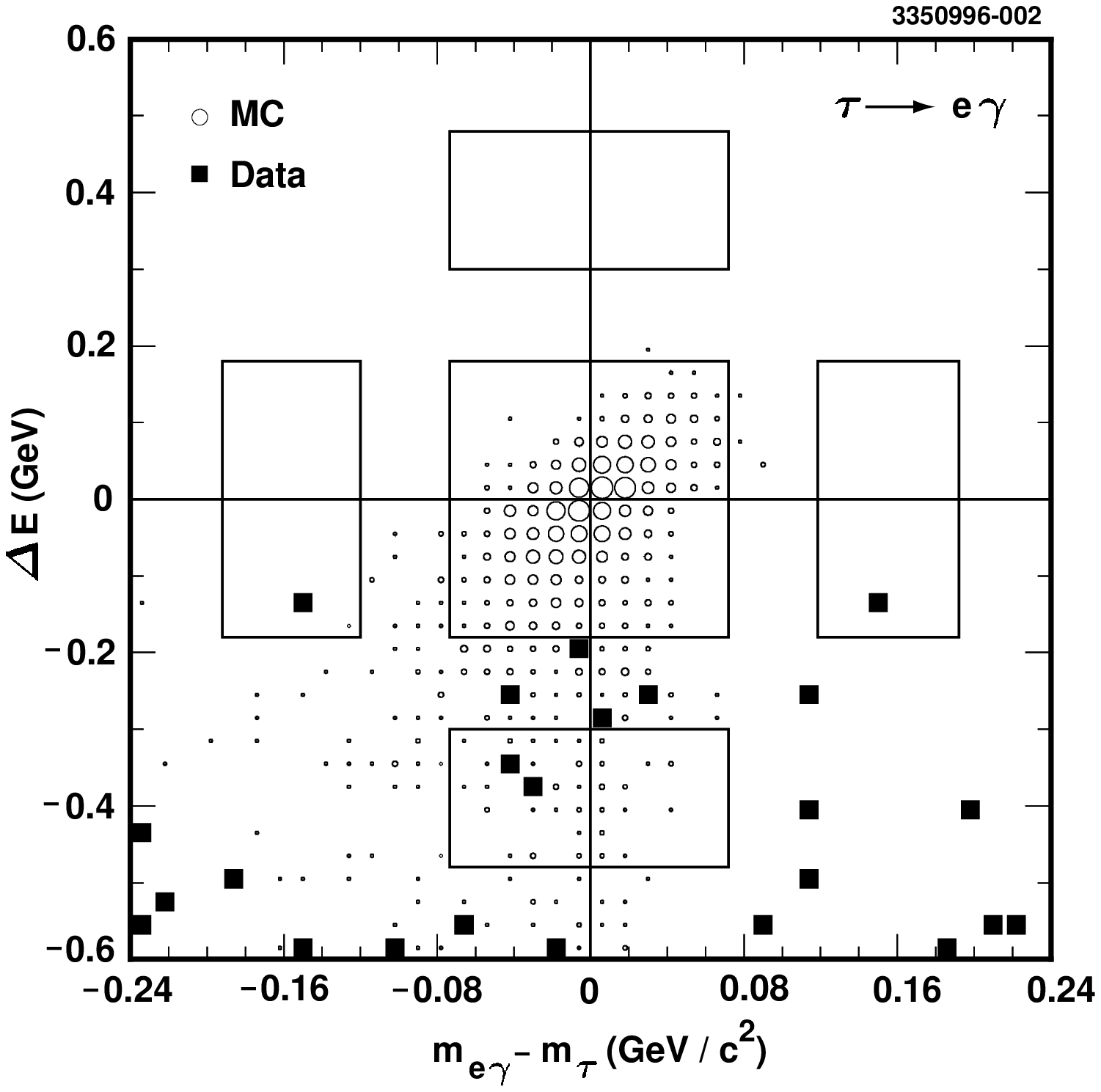,width=3.0in}}
\centerline{\psfig{figure=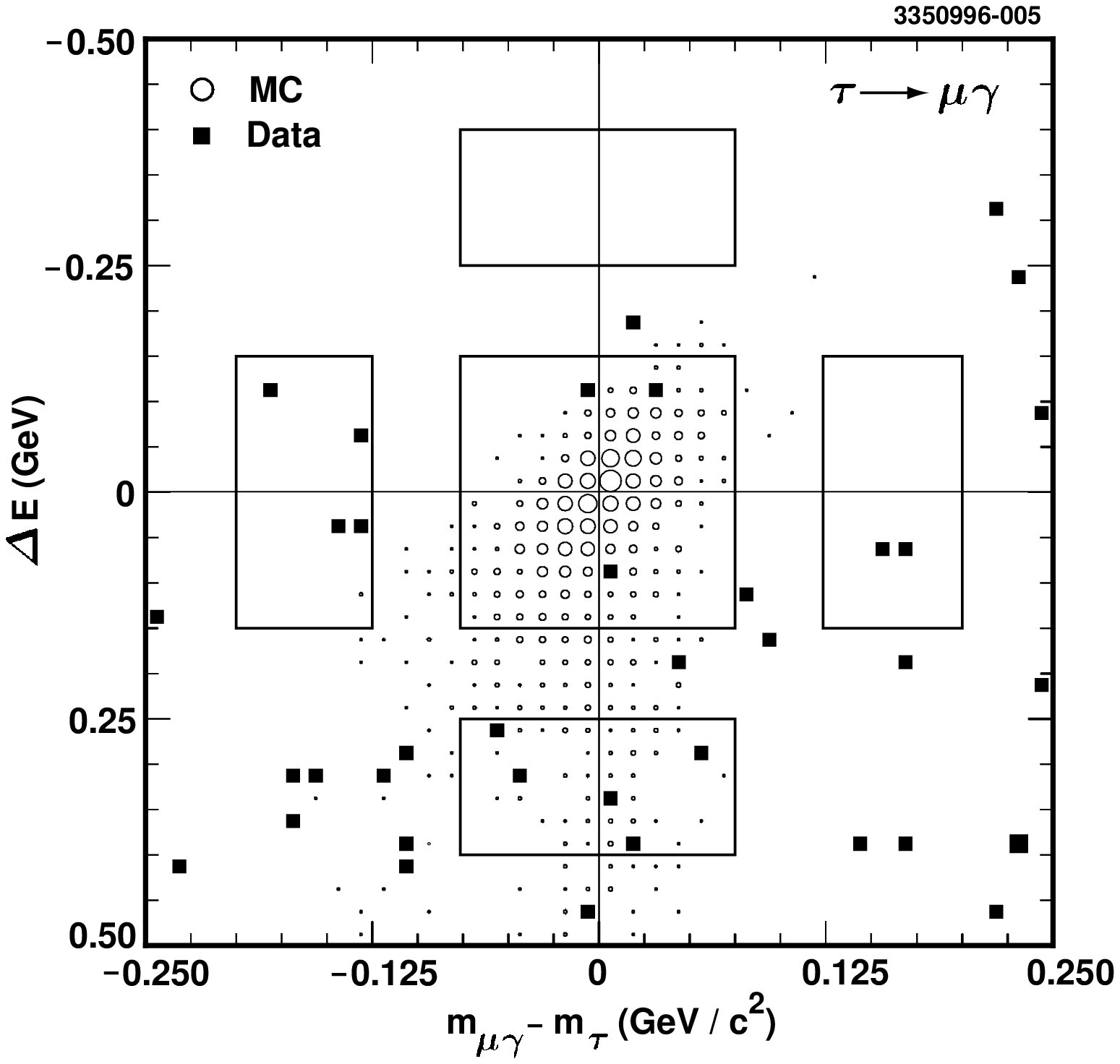,width=3.0in}}
\vspace{-0.5cm}
\caption{The total energy difference vs. the
invariant mass difference for (a) $\tauegamma$ and (b) $\taumugamma$.}
\label{fig:E_vs_m}
\end{figure}

The CLEO~II experiment has searched for the decays $\tauegamma$
and $\taumugamma$ \cite{Edwards}.
The experiment searches for the signal by examining the total energy
vs. the invariant mass of the lepton and photon in events that
survive the select criteria.
Figure~\ref{fig:E_vs_m} shows the scatter plots of
$\Delta E$ vs. $(m_{l\gamma} - m_\tau)$, where
$\Delta E = E_{l\gamma} - E_{beam}$ is the difference between
the measured total energy and the beam energy, $m_{l\gamma}$ is
the invariant mass of the lepton and photon, and $m_\tau$
is the mass of the $\tau$ lepton.
The signal region is defined as the $\pm 3\sigma$ region while the
sidebands are defined by the 5 to 8 $\sigma$ region.
In the signal region, there is no $e\gamma$ event while three $\mu\gamma$
events survive the selection criteria.
The background is estimated by extrapolating from the sideband regions,
assuming that the background is linear in this vicinity.
The estimated background is 2.0 events for $e\gamma$ and 5.5 events
for $\mu\gamma$.
The higher background for the $\mu\gamma$ analysis is due to the
looser selection criteria.
The observed events are consistent with background expectations.
Assuming Poisson statistics, the 90\% confidence level upper limits
on the branching ratios are:

\vspace{2mm}
\hspace{1cm} $\tauegamma < 2.7 \times 10^{-6}$

\vspace{2mm}
\hspace{1cm} $\taumugamma < 3.0 \times 10^{-6}$.

\vspace{2mm}
\noindent These represent considerable improvements over the
previous limits \cite{PDG},

\vspace{2mm}
\hspace{1cm} $\tauegamma < 1.2 \times 10^{-4}$\hspace{10mm} ARGUS

\vspace{2mm}
\hspace{1cm} $\taumugamma < 4.2 \times 10^{-6}$\hspace{10mm} CLEO

\vspace{2mm}
\noindent For comparison, the limit \cite{PDG} from $\mu$ decay is

\vspace{2mm}
\hspace{1cm} $\mu^- \ra e^-\gamma < 4.9 \times 10^{-11}$.

\vspace{2mm}
\noindent The new limit on $\taumugamma$ can be used to exclude
some parameter space such as the mass of certain supersymmetry
particles in some models \cite{Kelley,Hisano,Gomez}.

Experimenters have searched for neutrino-less decays in
37 modes as shown in Table~\ref{tab:limits}, including the
two radiative decays discussed above.
The CLEO~II experiment has contributed to 24 modes \cite{Bartelt2}
with limits of a few times $10^{-6}$, except $\tau^- \to
\mu^-\pi^-K^+$, $\mu^+\pi^-K^-$, and $e^-\bar K^{*0}(892)$
where the limits are at the $10^{-5}$ level.
These limits represents the most stringent limits, except
$\tau^- \to \mu^-\mu^+\mu^-$ where the ARGUS limit \cite{Albrecht}
remains the most stringent.
The limits on decays which have not been searched for by CLEO~II are
presently in the range of $10^{-3}$ to $10^{-5}$ and the experimenters
are urged to search for the decays.
There are also decays that have never been searched for:

\vspace{2mm}
\hspace{1cm} $\tau^- \to l^-\eta^\prime$, $l^-\omega$

\vspace{2mm}
\hspace{1cm} $\tau^- \to l^-\pi^0\pi^0$, $l^-\pi^0\eta$, $l^-\eta\eta$...

\vspace{2mm}
\hspace{1cm} $\tau^- \to l^-K^+K^-$, $l^-\bar K^0 K^0$...

\vspace{2mm}
\hspace{1cm} $\tau^- \to l^-\pi^0 K^0$, $l^-\eta K^0$, $l^-\eta^\prime K^0$...

\vspace{2mm}
\noindent Once again, the experimenters are urged to search for these decays.

\begin{table}[p]
\setlength{\tabcolsep}{1.5pc}
\caption{Upper limits on the branching ratios for neutrino-less decays.
All limits are at the 90\% confidence level, except the last two limits
which are at 95\% confidence level.
The decays $\tau^- \to \pi^-\gamma$ and $\pi^-\pi^0$ violate angular
momentum conservation.}
\label{tab:limits}
\begin{tabular*}{7.5cm}{@{}l@{\extracolsep{\fill}}cl}
\hline
Decay                    & $B$                    & Experiment\\ \hline
$e^-\gamma$              & $<2.7  \times 10^{-6}$ & CLEO   \\
$\mu^-\gamma$            & $<3.0  \times 10^{-6}$ & CLEO   \\
$e^-\pi^0$               & $<1.4  \times 10^{-4}$ & X'Ball \\
$\mu^-\pi^0$             & $<4.4  \times 10^{-5}$ & ARGUS  \\
$e^-K^0$                 & $<1.3  \times 10^{-3}$ & Mark II\\
$\mu^-K^0$               & $<1.0  \times 10^{-3}$ & Mark II\\
$e^-\eta$                & $<6.3  \times 10^{-5}$ & ARGUS  \\
$\mu^-\eta$              & $<7.3  \times 10^{-5}$ & ARGUS  \\
$e^-\rho^0$              & $<4.2  \times 10^{-6}$ & CLEO   \\
$\mu^-\rho^0$            & $<5.7  \times 10^{-6}$ & CLEO   \\
$e^-K^{*0}(892)$         & $<6.3  \times 10^{-6}$ & CLEO   \\
$\mu^-K^{*0}(892)$       & $<9.4  \times 10^{-6}$ & CLEO   \\
$\pi^-\gamma$            & $<2.8  \times 10^{-4}$ & ARGUS  \\
$\pi^-\pi^0$             & $<3.7  \times 10^{-4}$ & ARGUS  \\
$e^-e^+e^-$              & $<3.3  \times 10^{-6}$ & CLEO   \\
$e^-\mu^+\mu^-$          & $<3.6  \times 10^{-6}$ & CLEO   \\
$e^+\mu^-\mu^-$          & $<3.5  \times 10^{-6}$ & CLEO   \\
$\mu^-e^+e^-$            & $<3.4  \times 10^{-6}$ & CLEO   \\
$\mu^+e^-e^-$            & $<3.4  \times 10^{-6}$ & CLEO   \\
$\mu^-\mu^+\mu^-$        & $<1.9  \times 10^{-6}$ & ARGUS  \\
$e^-\pi^+\pi^-$          & $<4.4  \times 10^{-6}$ & CLEO   \\
$e^+\pi^-\pi^-$          & $<4.4  \times 10^{-6}$ & CLEO   \\
$\mu^-\pi^+\pi^-$        & $<7.4  \times 10^{-6}$ & CLEO   \\
$\mu^+\pi^-\pi^-$        & $<6.9  \times 10^{-6}$ & CLEO   \\
$e^-\pi^+K^-$            & $<7.7  \times 10^{-6}$ & CLEO   \\
$e^-\pi^-K^+$            & $<4.6  \times 10^{-6}$ & CLEO   \\
$e^+\pi^-K^-$            & $<4.5  \times 10^{-6}$ & CLEO   \\
$\mu^-\pi^+K^-$          & $<8.7  \times 10^{-6}$ & CLEO   \\
$\mu^-\pi^-K^+$          & $<1.5  \times 10^{-5}$ & CLEO   \\
$\mu^+\pi^-K^-$          & $<2.0  \times 10^{-5}$ & CLEO   \\
${\rm \bar p}\gamma$     & $<2.9  \times 10^{-4}$ & ARGUS  \\
${\rm \bar p}\pi^0$      & $<6.6  \times 10^{-4}$ & ARGUS  \\
${\rm \bar p}\eta$       & $<1.30 \times 10^{-3}$ & ARGUS  \\
$e^-\bar K^{*0}(892)$    & $<1.1  \times 10^{-5}$ & CLEO   \\
$\mu^-\bar K^{*0}(892)$  & $<8.7  \times 10^{-6}$ & CLEO   \\
$e^-{\rm light~boson}$   & $<2.7  \times 10^{-3}$ & ARGUS  \\
$\mu^-{\rm light~boson}$ & $<5    \times 10^{-3}$ & ARGUS  \\ \hline
\end{tabular*}
\end{table}

\begin{figure}[htb]
\centerline{\psfig{figure=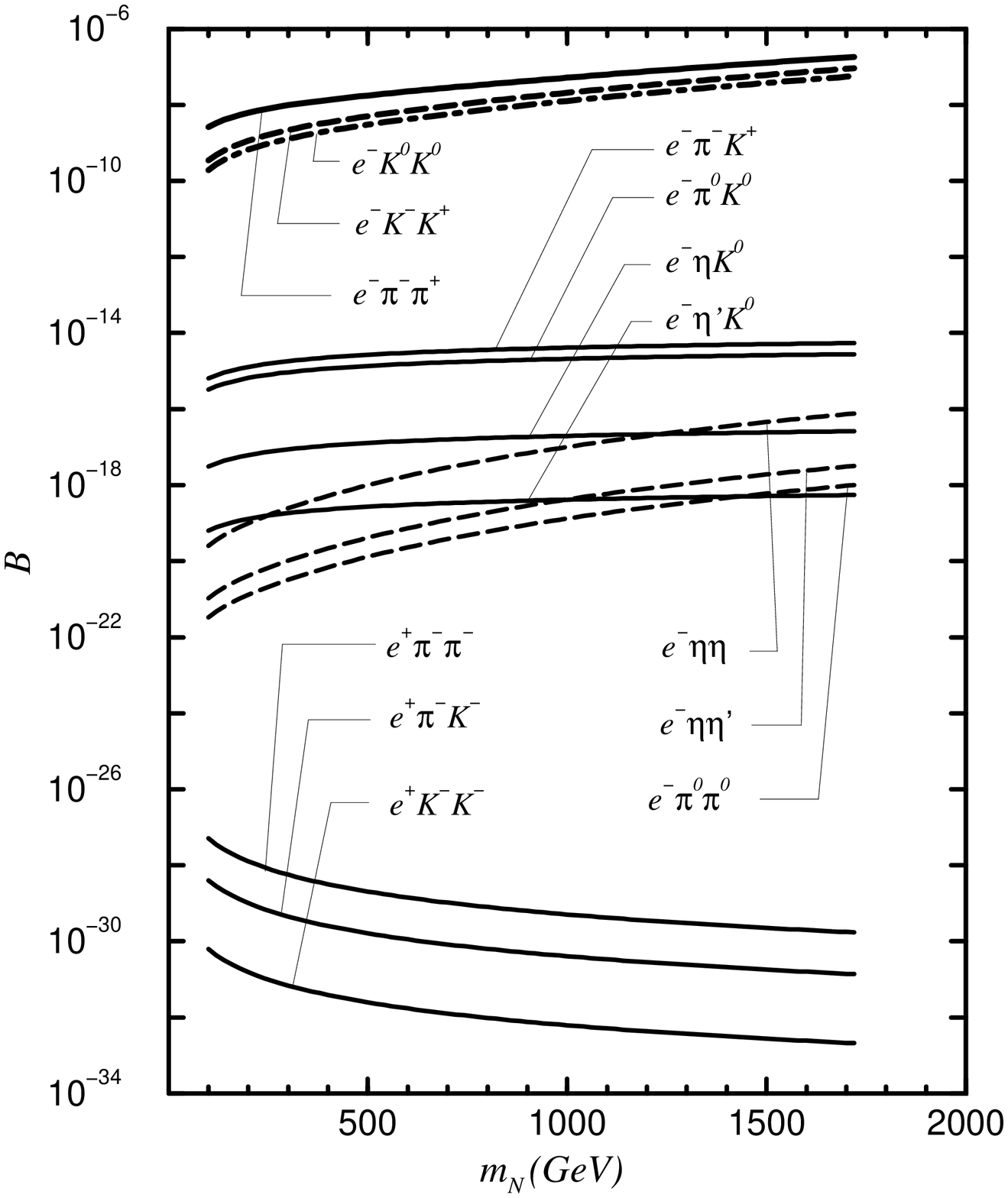,width=3.0in}}
\vspace{-0.5cm}
\caption{The predictions [31] for the decay branching ratios of
$\tau^-$ into one lepton and two mesons
as a function of the Majorana mass.}
\label{fig:Ilakov}
\end{figure}

Ilakovoc and collaborators \cite{Ilakovoc1,Ilakovoc2} have calculated
the rate for $\tau$ decay into three leptons and one lepton plus
one or two mesons using a GUT and superstring inspired model with
heavy neutral leptons.
The rates depend on the masses of the Majorana neutrinos, $M_{N_1}$ and
$M_{N_2}$, and the heavy-light neutrino mixings, $(s^{\nu_e}_L)^2$ and
$(s^{\nu_\tau}_L)^2$.
Some of the rates may be as large as $10^{-6}$, within the sensitivity of the
CLEO~II experiment.
Figure~\ref{fig:Ilakov} shows the dependence of the branching ratios
on the Majorana mass for the decay into one lepton and two mesons,
with $M_N = M_{N_1} = \frac{1}{3} M_{N_2}$, $(s^{\nu_e}_L)^2  = 0.01$ and
$(s^{\nu_\tau}_L)^2 = 0.05$.
The rates are largest for $\tau^- \to l^-\pi^+\pi^-$, $l^-K^+K^-$, and
$l^-\bar K^0 K^0$, which are enhanced by vector dominance, and
smallest for $\tau^- \to l^+\pi^-\pi^-$, $l^+\pi^-K^-$, and
$l^+ K^- K^-$, which proceed through tree level diagrams only.
There is no experimental limit on $\tau^- \to l^-K^+K^-$.

\section{Conclusion}

In conclusion, there are several new results on rare and forbidden
decays of the $\tau$ lepton.
The chiral anomaly decay $\taupietapio$ has been measured with good
precision and the result is somewhat higher than the prediction of CVC.
The second class current has been searched for in the decays $\taupieta$
and $\taupiomega$ and new upper limits have been set.
The Cabibbo-suppressed decay $\tauketa$ has been observed and the
measured branching ratio is consistent with the Standard Model expectation.
The internal conversion decay $\taueee$ has also been observed,
at a rate expected from the Standard Model.
There are also new upper limits
on the radiative decays $\tauegamma$ and $\taumugamma$.
In summary, we have reached a new level of sensitivity in $\tau$ physics.
We are now sensitive to branching ratios at the level of $10^{-6}$.
Unfortunately, there is no hint of physics beyond the Standard Model.

\section*{Acknowledgments}
This work was supported in part by the OJI program of
the U.S.~Department of Energy.
The author wishes to thank R.~Kass
for the careful reading of this manuscript and the organizers
for the wonderful meeting.

\end{document}

%% file: myphyx.tex
\def\etal    {\it et al.\rm}
\def\ee      {e^+e^-}

\def\gg      {\gamma\gamma}

\def\nutau   {\nu_\tau}
\def\nuebar  {\bar \nu_e}

\def\ra      {\rightarrow}

\def\taupieta        {\tau^- \ra \pi^-  \eta         \nutau}
\def\tauketa         {\tau^- \ra   K^-  \eta         \nutau}
\def\taupietapio     {\tau^- \ra \pi^-  \eta  \pi^0  \nutau}

\def\taupiomega      {\tau^- \ra \pi^-  \omega       \nutau}